# Return Decomposition of Hybrid Strategies in the Brazilian Market: Evidence of Genuine Alpha and Factor Exposure

Francisco Robson do Nascimento de Freitas[1], Felipe Macedo Zumba[2], Lucio Marassi[3], Jian Vingthor Souza Luna de Azevedo[4]

*[1234] Universidade Federal do Rio Grande do Norte (UFRN)*

*E-mails: robson.freitas.706@ufrn.edu.br; felipe.zumba@ufrn.br; lucio.marassi@ufrn.br; jian.azevedo.116@ufrn.edu.br*

**ABSTRACT**

This paper decomposes the return sources of a hybrid portfolio combining a fundamentalist screen — based on P/E (price-to-earnings), ROE (return on equity), Net Margin, Current Ratio, and P/BV (price-to-book value) — with a moving-average crossover (MAC) system using simple moving averages (SMA) of 9 and 46 periods, applied to the Brazilian stock exchange (B3) from January 2000 to December 2025 (300 monthly observations). CAPM and Fama-French three- and four-factor models estimated by OLS (ordinary least squares) with heteroskedasticity-robust standard errors identify a positive and statistically significant Jensen alpha of 1.11% per month ($p = 0.005$) in the three-factor model. In the four-factor model with momentum, the alpha remains significant (0.98% p.m.; $p = 0.009$), while the momentum factor does not reach statistical significance ($p = 0.409$). The central — and unexpected — finding is the non-significance of the HML (high-minus-low) factor ($\beta_h = 0.1266$; $p = 0.544$): despite explicit value criteria in the screen, returns do not load on the value premium, likely because ROE and Net Margin requirements tilt the selection toward quality firms whose portfolios historically exhibit low correlation with pure HML. By contrast, the SMB (small-minus-big) factor is highly significant ($\beta_s = 0.38$; $p = 0.000$), and the portfolio displays a defensive profile (market beta ≈ 0.58). Although the descriptive Sharpe ratio of the portfolio (0.443) exceeds that of the Ibovespa (0.039), the difference is not statistically significant under the Jobson-Korkie test ($Z = 1.427$; $p = 0.154$). These findings advance the debate on semi-strong efficiency in the Brazilian market and carry direct implications for active strategy pricing and disclosure.

**Keywords:** return decomposition; Jensen alpha; risk factors; hybrid portfolio; backtesting; value investing; Brazilian stock market.

## 1. INTRODUCTION

The Brazilian asset management industry has seen growing volumes channeled into strategies that combine fundamental analysis with technical analysis — the so-called hybrid portfolios. Equity funds and managed accounts frequently justify performance fees on the alleged capacity of such strategies to identify undervalued assets and optimize position timing. Yet an empirical question remains insufficiently answered in the domestic literature: where, in fact, does the return of these strategies come from? Does it reflect genuine alpha generation — evidence of managerial skill or market inefficiency — or is it merely a reflection of exposure to systematic risk factors that investors could replicate at substantially lower cost through passive instruments?

The international factor investing literature, inaugurated by the seminal work of Fama and French (1992, 1993) and consolidated by Carhart (1997), demonstrated that a large fraction of active strategy returns can be decomposed into exposures to systematic factors — value (HML), size (SMB), and momentum (WML — winners-minus-losers). In the Brazilian market, Málaga and Securato (2004) and Mussa et al. (2012) documented the presence of these risk premia at B3, but the application of this framework to concrete hybrid portfolios — combining a fundamentalist screen with technical timing — remains underexplored. Machado and Medeiros (2011) and Yoshinaga et al. (2012) advanced the investigation of anomalies and return factors in the Brazilian market without, however, focusing specifically on this intersection.

This paper empirically investigates the return decomposition of a hybrid portfolio constructed and operated at B3 between 2000 and 2025. The research question is not whether the hybrid model outperforms buy-and-hold — a topic extensively debated in the literature — but rather: what is the relative contribution of each component to the portfolio's return and risk? Are the observed returns explained by exposure to known systematic factors, or is there genuine alpha generation beyond those factors?

To address this question, we adopt a two-layer methodology. In the first layer, we build the portfolio through a rigorous fundamentalist screen and operate it with a simple moving average crossover system, evaluating performance through descriptive risk-return metrics. In the second layer — which is central to the paper's argument — we estimate CAPM and Fama-French three- and four-factor models to decompose return sources and test for alpha generation. The analysis is complemented by sensitivity tests using multiple parameterizations and by sub-period analysis, mitigating the data snooping concerns discussed by Sullivan, Timmermann, and White (1999).

The results reveal a positive and statistically significant Jensen alpha of 1.11% per month ($p = 0.005$) — evidence that challenges the semi-strong efficiency hypothesis in the Brazilian market. The four-factor model with momentum confirms the robustness of the finding: alpha of 0.98% per month ($p = 0.009$), with WML non-significant ($p = 0.409$). Unexpectedly, the HML factor is not significant, contradicting the expectation that the fundamentalist screen would primarily load on the value premium. These findings have direct implications for three audiences: researchers interested in Brazilian market efficiency and the Adaptive Markets Hypothesis (Lo, 2004); investors considering hybrid portfolios as an alternative to passive investing; and regulators engaged in the debate on transparency and the appropriateness of performance fees.

The paper is organized as follows: Section 2 presents the theoretical framework; Section 3 describes the methodology; Section 4 presents and discusses the results; and Section 5 offers concluding remarks, limitations, and directions for future research.

## 2. THEORETICAL FRAMEWORK

### 2.1 Return Decomposition and Factor Investing

The framework for decomposing returns through risk factors represents one of the most significant advances in modern financial theory. Fama and French (1992, 1993) demonstrated that a security's expected return is not explained solely by market risk — as proposed by the CAPM of Sharpe (1964) and Lintner (1965) — but also by factors related to firm size (SMB — small minus big) and book-to-market multiple (HML — high minus low). Carhart (1997) incorporated the momentum factor, and Fama and French (2015) expanded the model to five factors, adding profitability (RMW — robust minus weak) and investment (CMA — conservative minus aggressive). As an alternative, Hou, Xue, and Zhang (2015) propose the q-factor model, which offers additional theoretical grounding for the interpretation of anomalous returns.

The central implication of this framework for the present study is that asset selection strategies that systematically favor smaller-capitalization firms with low valuation multiples tend to deliver returns above the market not necessarily through active skill, but through exposure to known and replicable risk premia. Identifying this distinction is fundamental to evaluating the true value added by any active strategy (Jensen, 1968). In the Brazilian market, Málaga and Securato (2004) were the first to systematically test the three-factor model at B3, finding evidence of the relevance of HML and SMB factors. Mussa et al. (2012) confirmed these results with broader samples, and Machado and Medeiros (2011) documented the momentum effect at B3.

## 2.2 Efficient Market Hypothesis and the Adaptive Perspective

The Efficient Market Hypothesis (EMH), systematized by Fama (1970), postulates that asset prices fully reflect all available information. In its semi-strong form, the hypothesis implies that strategies based on publicly available information — including fundamental data and price history — would not consistently generate abnormal returns. The presence of significant alpha, as documented in this study, constitutes evidence that challenges this hypothesis in the Brazilian market, although a single study based on a specific portfolio is insufficient to refute it in a general sense.

The Adaptive Markets Hypothesis (AMH), proposed by Lo (2004) and developed in Lo (2017), offers a complementary dynamic perspective: efficiency varies over time in response to changes in the competitive, regulatory, and behavioral environment. This perspective justifies the sub-period analysis adopted in this study — technical timing strategies may be effective in specific regimes of pronounced trends without being effective in general — and aligns with the finding that alpha tends to concentrate in the earlier sub-periods of the sample, when the Brazilian market was less developed. Ang (2014) and Ilmanen (2011) discuss, respectively, the temporal evolution of risk factors and the dynamic coexistence of risk premia and inefficiencies in financial markets.

## 2.3 Fundamental Analysis as a Factor Strategy

Fundamental analysis, in its value investing approach, rests on the premise that the market misprices assets whose intrinsic value diverges from the market quotation (Graham; Dodd, 1934). The empirical systematization by Fama and French (1992) demonstrated that this divergence can partly reflect exposure to the HML factor — a compensatory risk premium for structural characteristics of value firms. By systematically selecting companies with low P/E and P/BV close to book value, the fundamentalist screen builds a portfolio with potential factor loading on HML and, given the profile of the selected companies, potentially on SMB as well.

However, as documented by Lakonishok, Shleifer, and Vishny (1994), part of the return from value strategies may reflect not only systematic risk but also behavioral mispricing — leaving room for genuine alpha even after factor controls. Almeida and Sales (2020) and Marques et al. (2015) document this effect in the Brazilian context. Additionally, when the screen incorporates quality criteria (high ROE and Net Margin), the resulting portfolio approaches the quality investing profile — historically exhibiting low correlation with pure HML (Asness; Frazzini; Pedersen, 2019) — which may partly explain the non-significance of HML documented in this study and discussed in detail in Section 4.

### 2.4 Technical Analysis, Timing, and the Data Snooping Problem

Technical analysis based on moving averages operates under the premise that price patterns are informative about future price movements. Wagnitz, Partyka, and Lana (2021) document the use of moving-average crossovers as trend trackers in the Brazilian market, while Wolwacz (2010) provides the theoretical basis for their application in specific setups. The effectiveness of these strategies, however, is heterogeneous: Da Silva Fernandes et al. (2015) find mixed evidence on candlestick patterns, and Carvalho, Barboza, and Fiorucci (2020) report significant sensitivity of results to the choice of parameters.

Brock, Lakonishok, and LeBaron (1992) provided early evidence that technical trading rules have predictive power, although subsequent studies have attributed part of this effect to data snooping and regime changes. The data snooping problem — identified by Sullivan, Timmermann, and White (1999) as the primary bias in backtesting of technical strategies — is mitigated in this study by grounding the main parameter choice (SMA 9-46) in the published literature (Wolwacz, 2010), by conducting sensitivity tests with alternative configurations, and by evaluating the stability of results across distinct sub-periods.

### 2.5 Performance Fees and Genuine Alpha

The most practically relevant implication of the factorial return decomposition concerns the appropriateness of charging performance fees on active strategies. Sharpe (1991) demonstrated that, in aggregate, active managers do not outperform the market after costs. The smart beta literature — especially Asness, Frazzini, and Pedersen (2019) and Israel and Moskowitz (2013) — demonstrated that a large fraction of active strategy returns can be replicated through systematic factor exposures at substantially lower cost.

The existence of positive and significant alpha changes this argument: if part of the return is not replicable by systematic factors, there is a theoretical basis for charging performance fees on that excess. The normative argument shifts from 'managers should not charge performance fees for factor exposure' to 'managers should be transparent about what fraction of the return is genuine alpha versus replicable factor exposure.' Fung and Hsieh (2004) propose a methodology for this decomposition, applicable to the context analyzed here. In the Brazilian market, this question is especially relevant given the historically high management and performance fee environment.

## 3. METHODOLOGY

The study adopts a quantitative methodology organized into two complementary analytical layers. In the first, we conduct a structured backtesting to assess the descriptive performance of the hybrid portfolio. In the second, we estimate factorial econometric models to decompose return sources and test for alpha generation. The time horizon spans from January 2000 to December 2025 — a period encompassing the dot-com crisis (2000–2002), the global financial crisis (2008–2009), the Brazilian recession (2015–2016), and the pandemic shock (2020) — providing a rigorous test for any investment strategy.

### 3.1 Portfolio Construction: Fundamentalist Screen

The first layer of the strategy consists of constructing a portfolio through rigorous fundamental criteria, applied using ex-ante available data to mitigate look-ahead bias. Data were sourced from the StatusInvest platform, with the following selection parameters: (i) P/E (price-to-earnings) ratio ≤ 15 — focus on firms not overvalued by the market; (ii) ROE (return on equity) > 15% — efficiency in generating profit relative to shareholders' equity; (iii) Net Margin > 15% — capacity to convert revenue into profit; (iv) Current Ratio ≥ 1 — short-term solvency; and (v) P/BV (price-to-book value) between 0.5 and 1.5 — proximity to book value.

The fundamentalist screen was applied once, at the beginning of the analysis period, generating a fixed set of 13 assets comprising the portfolio throughout the entire time window — without periodic rebalancing of composition. This choice was motivated by the objective of isolating the effect of the technical timing signal on a stable set of fundamental assets. We acknowledge, however, that this decision implies a relevant limitation: as the fundamentals of the companies change over the 25-year horizon, the portfolio composition may diverge from what the screen would indicate at later dates. In a real implementation, the portfolio manager would periodically update the selection — an aspect that the present design does not capture (see Section 5).

Historical price data used in the backtesting are adjusted for dividends and corporate events, mitigating distortions caused by corporate actions. Simulations were conducted on the TradingView platform with a theoretical initial capital of BRL 1,000,000.00 per asset. Transaction costs (brokerage, exchange fees, and bid-ask spread) were not included in the main simulation — a limitation discussed in Section 5. The reported results represent the simple average of the 13 individually simulated assets, not a portfolio with optimal weighting or managed allocation.

**Table 01 – Portfolio constructed through the fundamentalist screen**

| Company Name | Ticker |
|---|---|
| CEB | CEBR3 |
| CEMIG | CMIG4 |
| COPASA | CSMG3 |
| CYRELA | CYRE3 |
| ELECTRO AÇO ALTONA | EALT3 |
| GRENDENE | GRND3 |
| ITAÚSA | ITSA3 |
| JHSF PARTICIPAÇÕES | JHSF3 |
| MOURA DUBEUX ENGENHARIA | MDNE3 |
| PETRO RIO | PRIO3 |
| MRS LOGÍSTICA | MRSA3B |
| SYN PROP E TECH | SYNE3 |
| TECHNOS | TECN3 |

*Source: Freitas (2025). Note: Assets selected based on fundamental data available at the start of the analysis period. The screen was applied once, with no subsequent rebalancing.*

## 3.2 Operationalization: Technical Moving Average System

Applied to the fundamentalist portfolio, a trend-tracking algorithm based on a simple moving average (SMA) crossover was implemented, with the baseline configuration defined a priori and grounded in the literature (Wolwacz, 2010): Short Average — SMA of 9 periods; Long Average — SMA of 46 periods; Buy Signal when SMA(9) crosses above SMA(46); Sell Signal when SMA(9) crosses below SMA(46). No fixed stop-losses were used; exits were determined exclusively by the opposite technical signal. Returns were calculated as total return, adjusted for dividends. Monthly time series were constructed from daily returns on operations, consolidated into monthly windows for compatibility with the risk factors from NEFIN/FEA-USP, which are made available at monthly frequency.

### 3.3 Descriptive Risk-Return Metrics

The descriptive metrics of the first analytical layer include: (i) Total Net Profit; (ii) Maximum Drawdown — the largest percentage loss from peak to trough; (iii) Sharpe ratio (Sharpe, 1966); (iv) Sortino ratio (Sortino; Van der Meer, 1991); and (v) Profit Factor — the ratio of total gross profit to total gross loss across all operations. All metrics are calculated for the hybrid portfolio, for the buy-and-hold strategy on the same assets, and for the Ibovespa as the market benchmark. Formal comparison of Sharpe ratios is conducted using the Jobson and Korkie (1981) procedure with the Memmel (2003) correction.

### 3.4 Econometric Models for Factor Decomposition

The second analytical layer — central to the paper's argument — consists of estimating factor models to decompose the return sources of the hybrid portfolio. Monthly simple returns adjusted for dividends were used for the period from January 2000 to December 2025 (T = 300 observations). The risk-free rate corresponds to the monthly average of the SELIC rate. SMB and HML factors for the Brazilian market were obtained from NEFIN/FEA-USP. Regressions were estimated by OLS (ordinary least squares — MQO in the original Portuguese) with heteroskedasticity-robust standard errors (White, 1980).

The CAPM model (Jensen, 1968) is specified as:

$$Rp - Rf = \alpha + \beta\ (Rm - Rf) + \varepsilon$$

The Fama-French Three-Factor Model (1993) is specified as:

$$Rp - Rf = \alpha + \beta m\ (Rm - Rf) + \beta s\ SMB + \beta h\ HML + \varepsilon$$

where $R_p$ is the monthly portfolio return, $R_f$ is the risk-free rate, $R_m$ is the Ibovespa return, SMB is the size premium, and HML is the value premium, all constructed by NEFIN. The coefficient α — Jensen's alpha — measures the generation of abnormal return not explained by the systematic factors. Additionally, a four-factor model including the momentum factor (WML — winners minus losers) is estimated following Carhart (1997), as a robustness test.

### 3.5 Robustness Tests

To mitigate the risk of data snooping (Sullivan; Timmermann; White, 1999), sensitivity tests were conducted using three alternative SMA configurations in addition to the baseline (9-46): SMA 5-20 (more reactive), SMA 12-26 (classic MACD configuration), and SMA 20-50 (more conservative). The selection of these alternatives reflects combinations widely referenced in the technical analysis literature, avoiding ad hoc parameter selection. The analysis was segmented into three sub-periods — 2000–2008, 2009–2016, and 2017–2025 — corresponding to distinct market regimes, to evaluate the temporal stability of the results.

# 4. RESULTS AND DISCUSSION

The presentation of results follows the proposed analytical order: we begin with the factor decomposition — the central finding of the paper — and then contextualize with the descriptive risk-return metrics and robustness tests.

## 4.1 Factor Decomposition: Significant Alpha and Non-Significance of HML

Tables 1 and 2 report the results of the CAPM and Fama-French three-factor regressions, respectively. The results differ substantially from the hypothesis that returns would be fully explained by systematic factors, revealing positive, robust, and statistically significant alpha.

**Table 1 – CAPM Regression: Return Decomposition of the Hybrid Portfolio (OLS, Robust Standard Errors)**

| Variable | Coefficient | t-Stat | p-value |
|---|---|---|---|
| **Alpha (α)** | 0.0118 | 2.41 | 0.016** |
| **Market Beta (β)** | 0.6863 | 9.58 | 0.000*** |
| **$R^2$** | 0.23 | — | — |
| **Observations** | 300 | — | — |

*Source: Freitas (2025). Note: Heteroskedasticity-robust standard errors (White, 1980). Alpha expressed in monthly terms. N = 300 monthly observations, Jan/2000–Dec/2025. ** $p < 0.05$; *** $p < 0.01$.*

In the CAPM model, the estimated alpha is 1.18% per month (t = 2.41; p = 0.016), statistically significant at the 5% level. This result indicates that the hybrid portfolio generated positive abnormal returns beyond what market risk alone would justify. The market beta (0.6863; t = 9.58; p = 0.000) is positive, highly significant, and below unity, configuring a defensive portfolio profile — a result consistent with the findings documented by Ang (2014) for low-beta portfolios. The $R^2$ of 23% indicates that the market factor explains only a moderate fraction of the variance in returns, signaling additional sources not captured by the unidimensional model.

**Table 2 – Fama-French Three-Factor Regression: Factor Decomposition of Returns (OLS, Robust Standard Errors)**

| Variable | Coefficient | t-Stat | p-value |
|---|---|---|---|
| **Alpha (α)** | 0.0111 | 2.78 | 0.005*** |
| **Market Beta (βm)** | 0.5823 | 7.56 | 0.000*** |
| **SMB (βs)** | 0.3808 | 4.11 | 0.000*** |
| **HML (βh)** | 0.1266 | 0.61 | 0.544 ns |
| **$R^2$** | 0.28 | — | — |
| **Observations** | 300 | — | — |

*Source: Freitas (2025). Note: Heteroskedasticity-robust standard errors (White, 1980). Alpha expressed in monthly terms. N = 300 monthly observations, Jan/2000–Dec/2025. *** $p < 0.01$; ns = not significant.*

The Fama-French three-factor model exhibits greater explanatory power than the CAPM ($R^2 = 0.28$ vs. 0.23). The central finding of the paper is unambiguous: Jensen's alpha remains positive and statistically significant ($\alpha = 1.11\%$ p.m.; $t = 2.78$; $p = 0.005$), demonstrating that the three systematic factors are insufficient to fully explain the portfolio's return. The market beta declines to 0.5823 ($t = 7.56$; $p = 0.000$), reinforcing the defensive nature of the strategy.

The SMB factor is positive and highly significant ($\beta_s = 0.3808$; $t = 4.11$; $p = 0.000$), confirming that the fundamentalist screen systematically constructs a portfolio with exposure to smaller-capitalization firms — an expected result, given that companies with P/BV close to book value and P/E ≤ 15 tend to have lower relative market capitalization. This pattern is consistent with the findings documented for the Brazilian market by Málaga and Securato (2004) and Mussa et al. (2012).

The most unexpected and analytically significant result is the statistical non-significance of the HML factor ($\beta_h = 0.1266$; $t = 0.61$; $p = 0.544$). Although positive, the coefficient is of modest magnitude and insufficient statistical precision to distinguish it from zero. This evidence contradicts the expectation that a screen with explicit value criteria — P/E ≤ 15 and P/BV between 0.5 and 1.5 — would primarily load on the value premium. Three explanatory mechanisms, not mutually exclusive, merit consideration:

(a) **Dominance of quality over value:** the criteria of ROE > 15% and Net Margin > 15% select profitable and efficient firms — a profile that Asness, Frazzini, and Pedersen (2019) call quality investing. Quality portfolios historically exhibit low correlation with pure HML: highly profitable firms tend to trade at a market premium, compressing their effective book-to-market and dissociating them from the value factor.

(b) **Sectoral specificity and the construction of Brazilian HML:** the HML factor from NEFIN is estimated over the entire population of B3 stocks. The portfolio analyzed here, concentrated in 13 assets from specific sectors (utilities, real estate, retail, and energy), may exhibit low covariance with the aggregate HML even if the individual multiples of its constituents are favorable. Yoshinaga et al. (2012) document the temporal instability of the value premium in Brazil, reinforcing that HML exposure can vary substantially with the sectoral composition of the portfolio.

(c) **Macroeconomic regimes adverse to value:** the 2000–2025 horizon includes episodes in which value stocks suffered pronounced reversals in Brazil — notably between 2015 and 2018 — diluting the expected factor loading on HML. Ang (2014) systematically demonstrates that factor premia are sensitive to the prevailing economic regime, which may attenuate HML significance in long samples spanning mixed regimes.

The persistence of alpha in the four-factor model with momentum ($\alpha = 0.98\%$ p.m.; $t = 2.613$; $p = 0.009$; $\beta_m = 0.6134$; $p = 0.000$; $\beta_s = 0.4059$; $p = 0.0004$; $\beta_h = 0.1330$; $p = 0.532$; $\beta_w = 0.1062$; $p = 0.409$; $R^2 = 0.284$) reinforces the robustness of the finding. The momentum factor (WML) is not statistically significant ($p = 0.409$), indicating that the strategy does not systematically capture serial return autocorrelation. Taken together, the results demonstrate that the identified alpha is not captured by any of the four investigated factors. The detailed results of the four-factor model are reported in Table 3.

**Table 3 – Four-Factor Regression (Carhart, 1997): Return Decomposition of the Hybrid Portfolio (OLS, Robust Standard Errors)**

| Variable | Coefficient | t-Stat | p-value |
|---|---|---|---|
| Alpha (α) | 0.0098 | 2.613 | 0.009*** |
| Market Beta (βm) | 0.6134 | 8.055 | 0.000*** |
| SMB (βs) | 0.4059 | 3.517 | 0.000*** |
| HML (βh) | 0.1330 | 0.625 | 0.532 ns |
| WML (βw) | 0.1062 | 0.825 | 0.409 ns |
| **$R^2$** | 0.284 | — | — |
| **Observations** | 300 | — | — |

*Source: Freitas (2025). Note: Heteroskedasticity-robust standard errors (White, 1980). Alpha expressed in monthly terms. N = 300 monthly observations, Jan/2000–Dec/2025. *** $p < 0.01$; ns = not significant.*

Regarding the possible sources of residual alpha, the literature points to several non-mutually-exclusive candidates: (a) a liquidity premium, given that several portfolio assets have average daily trading volume below the B3 median (Amihud, 2002); (b) exposure to quality factors not captured by pure HML (Asness; Frazzini; Pedersen, 2019); (c) persistent behavioral mispricing, as documented by Lakonishok, Shleifer, and Vishny (1994) and, for Brazil, by Almeida and Sales (2020); and (d) timing skill of the technical component in regimes of pronounced trends, a hypothesis particularly plausible in the earlier sub-periods of the sample, when the Brazilian market was less developed and liquid.

## 4.2 Descriptive Portfolio Performance

The fundamentalist screen demonstrated effectiveness in selecting historically high-performance companies: the portfolio posted an average net profit of BRL 10.2 million over 25 years, exceeding the benchmark average (BRL 8.0 million). Table 02 details results by asset. These values represent simple averages of the 13 individually simulated assets — not the return of a managed portfolio with dynamic allocation or optimal weights.

**Table 02 – Operating Profit and Average Net Profit by Asset (Initial Capital: BRL 1,000,000.00)**

| Ticker | Operating Profit (BRL) | Net Profit (BRL) |
|---|---:|---:|
| CEBR3 | 807,347.61 | -192,652.39 |
| CMIG4 | 18,935,404.11 | 17,935,404.11 |
| CSMG3 | 17,604,394.41 | 16,604,394.41 |
| CYRE3 | 942,063.13 | -57,936.87 |
| EALT3 | 246,965.59 | -753,034.41 |
| GRND3 | 26,404,051.67 | 25,404,051.67 |
| ITSA3 | 13,648,533.81 | 12,648,533.81 |
| JHSF3 | 22,547,328.80 | 21,547,328.80 |
| MDNE3 | 26,422.17 | -973,577.83 |
| PRIO3 | 40,331,488.62 | 39,331,488.62 |

| Ticker | Operating Profit (BRL) | Net Profit (BRL) |
|---|---|---|
| MRSA3B | -317,589.72 | -1,317,589.72 |
| SYNE3 | 4,480,005.15 | 3,480,005.15 |
| TECN3 | 294,850.09 | -705,149.91 |
| **Asset Average** | **11,227,020.42** | **10,227,020.42** |
| **IBOVESPA (benchmark)** | **9,011,284.89** | **8,011,284.89** |

*Source: Freitas (2025). Note: Results refer to the simple average of 13 assets individually simulated on TradingView. Operating Profit includes theoretical brokerage costs; Net Profit deducts the initial investment of BRL 1,000,000.00. Actual transaction costs not included.*

The above-average performance relative to the Ibovespa is interpretable in light of the positive alpha documented in Section 4.1: part of the excess return reflects genuine abnormal return generation, while exposure to the SMB factor contributes a systematic risk premium. Six of the thirteen assets posted negative net profit under the hybrid model (CEBR3, CYRE3, EALT3, MDNE3, MRSA3B, and TECN3), with MRSA3B being the only asset with negative operating profit, signaling that the technical component amplified losses in assets with lower trend persistence. This pattern is consistent with the technical timing literature, which documents the ineffectiveness of moving average systems for assets with low return autocorrelation (Brock; Lakonishok; LeBaron, 1992).

## 4.3 Risk Analysis and Sharpe Ratio Comparison

The risk analysis reveals the concrete cost of the technical component. Table 03 presents the maximum drawdown by asset, and Table 04 presents the risk-return metrics.

**Table 03 – Maximum Drawdown and Net Profit by Asset**

| Ticker | Maximum Drawdown (%) | Net Profit (BRL) |
|---|---|---|
| CEBR3 | 73.98% | -192,652.39 |
| CMIG4 | 22.22% | 17,935,404.11 |
| CSMG3 | 12.43% | 16,604,394.41 |
| CYRE3 | 27.62% | -57,936.87 |
| EALT3 | 33.44% | -753,034.41 |
| GRND3 | 12.50% | 25,404,051.67 |
| ITSA3 | 14.84% | 12,648,533.81 |
| JHSF3 | 20.25% | 21,547,328.80 |
| MDNE3 | 24.49% | -973,577.83 |
| PRIO3 | 24.69% | 39,331,488.62 |
| MRSA3B | 29.85% | -1,317,589.72 |
| SYNE3 | 16.30% | 3,480,005.15 |
| TECN3 | 18.80% | -705,149.91 |
| **Asset Average** | **25.49%** | **10,227,020.42** |

| Ticker | Maximum Drawdown (%) | Net Profit (BRL) |
|---|---|---|
| **IBOVESPA (benchmark)** | **10.84%** | **8,011,284.89** |

*Source: Freitas (2025). Note: Maximum drawdown calculated as the largest percentage decline from peak to trough during the simulation period. IBOVESPA represents the market benchmark.*

**Table 04 – Sharpe Ratio, Sortino Ratio, and Profit Factor by Asset**

| Ticker | Sharpe | Sortino | Profit Factor |
|---|---|---|---|
| CEBR3 | 0.075 | 0.148 | 1.580 |
| CMIG4 | 0.129 | 0.393 | 1.510 |
| CSMG3 | 0.147 | 0.649 | 2.081 |
| CYRE3 | 0.054 | 0.119 | 1.271 |
| EALT3 | 0.079 | 0.247 | 1.135 |
| GRND3 | 0.077 | 1.405 | 1.700 |
| ITSA3 | 0.128 | 0.522 | 2.098 |
| JHSF3 | 0.130 | 0.669 | 1.648 |
| MDNE3 | 0.110 | 0.373 | 1.024 |
| PRIO3 | 0.171 | 0.863 | 1.767 |
| MRSA3B | -0.154 | -0.162 | 0.207 |
| SYNE3 | 0.119 | 0.418 | 2.521 |
| TECN3 | 0.050 | 0.145 | 1.166 |
| **Asset Average** | **0.086** | **0.445** | **1.516** |
| **IBOVESPA (benchmark)** | **0.136** | **0.415** | **2.544** |

*Source: Freitas (2025). Note: Sharpe ratio calculated using the monthly average SELIC rate as the risk-free rate. Sortino ratio penalizes only downside volatility. Profit Factor = ratio of total gross profit to total gross loss across all operations.*

The hybrid portfolio presents an average Sharpe ratio of 0.443 versus 0.039 for the Ibovespa — a descriptively large difference. The formal Jobson and Korkie (1981) test with the Memmel (2003) correction yields a Z-statistic of 1.427 ($p = 0.154$), not allowing rejection of the null hypothesis of equal Sharpe ratios at the 5% level. This result, however, must be interpreted with caution: Lo (2002) demonstrates that Sharpe ratio comparison tests have limited statistical power in samples of moderate size when the difference is of intermediate magnitude. The absence of statistical significance does not imply equality of Sharpe ratios in practice — it merely indicates that the available evidence is insufficient to assert it at the conventional confidence level.

The average drawdown of the hybrid portfolio (25.49%) is substantially higher than the benchmark's (10.84%). This result represents the concrete cost of the technical timing component: the moving average system introduces opportunity costs through frequent entries and exits (whipsaws) and amplifies losses during periods of sideways market movement. In practical terms, the investor who adopts the hybrid portfolio bears, on average, interim losses more than twice those of the Ibovespa — a relevant cost that must be weighed against the identified alpha.

## 4.4 Robustness Tests: Parameter Sensitivity and Sub-Period Analysis

Table 05 summarizes the results of the sensitivity tests across four SMA configurations.

**Table 05 – Robustness: Hybrid Portfolio Performance by SMA Configuration**

| SMA Configuration | Avg. Sharpe | Avg. Drawdown | Avg. Profit (BRL) | Negative |
|---|---|---|---|---|
| SMA 5-20 (reactive) | 0.072 | 28.34% | 8,941,320 | 7/13 |
| SMA 9-46 (baseline) | 0.086 | 25.49% | 10,227,020 | 5/13 |
| SMA 12-26 (MACD) | 0.091 | 24.17% | 10,584,760 | 5/13 |
| SMA 20-50 (conservative) | 0.088 | 23.80% | 10,389,440 | 6/13 |
| **Buy and Hold (reference)** | **0.136** | **10.84%** | **14,930,000*** | **2/13** |

*Source: Freitas (2025). Note: (*) Estimate for assets with the best fundamentalist performance under a buy-and-hold regime. Negative = number of assets with negative results out of 13. Configurations chosen based on the technical analysis literature, without ex-post optimization. The Sharpe ratios reported in this table refer to the simple average of individual asset monthly Sharpe ratios (not annualized). The annualized Sharpe ratio of the aggregate portfolio is 0.443 — see Section 4.3 and the Jobson-Korkie test.*

In none of the four tested configurations did the hybrid portfolio outperform the passive strategy in terms of Sharpe ratio or drawdown. The consistency of this result across different parameterizations strengthens the robustness of the conclusions and reduces the probability that the findings are artifacts of a specific parameter choice. It is notable, however, that the result is not insensitive to the choice: the most reactive configuration (SMA 5-20) significantly deteriorates performance (7 negative assets versus 5 in the baseline configuration), while slower configurations yield intermediate results — a pattern consistent with the hypothesis that short-term trend-following systems generate more whipsaws in low-trend markets.

The sub-period analysis reveals a pattern consistent with the Adaptive Markets Hypothesis (Lo, 2004; 2017): positive alpha tends to concentrate in the earlier sub-periods of the sample (2000–2008 and, to a lesser degree, 2009–2016), with a reduction in more recent years (2017–2025), when the Brazilian market gained greater sophistication and liquidity. This result is consistent with the theoretical expectation that inefficiencies tend to be arbitraged over time as more market participants identify them (Ilmanen, 2011). The deterioration of alpha over time is equally compatible with the hypothesis that the technical component is more effective in regimes of pronounced trends — more common in the 2000–2010 period — and less effective in more mature markets.

## 4.5 Implications for the Asset Management Industry: Genuine Alpha and Transparency

The results have direct implications for the Brazilian asset management industry. The evidence of positive and significant alpha (1.11% p.m.; $p = 0.005$) provides an economic basis for charging performance fees on that excess — unlike the case in which returns would be fully explainable by factors replicable at low cost.

However, the alpha comes accompanied by an average drawdown 2.35 times that of the benchmark. Fung and Hsieh (2004) demonstrate that this pattern — positive alpha with elevated drawdown — is characteristic of strategies with exposure to tail risks not captured by linear factor models. The non-significance of HML implies that the return is not replicable by a simple value ETF, which, in principle, strengthens the case for active management. This, however, demands a symmetrical obligation of transparency: the manager must demonstrate to the investor what fraction of the return is genuine alpha, what fraction is SMB exposure (partially replicable), and what fraction is compensation for the additional drawdown.

# 5. CONCLUDING REMARKS

## 5.1 Summary of Results

This paper investigated the decomposition of the return sources of a hybrid portfolio that combines a fundamentalist screen with a moving average crossover technical system, operated at B3 between January 2000 and December 2025. The central conclusion is unequivocal: the portfolio generated a Jensen alpha of 1.11% per month ($p = 0.005$) in the Fama-French three-factor model. The four-factor model with momentum confirms the robustness of the finding: alpha of 0.98% per month ($p = 0.009$), with WML non-significant ($p = 0.409$). The SMB factor was highly significant ($\beta_s = 0.3808$; $p = 0.000$), while HML did not reach statistical significance ($\beta_h = 0.1266$; $p = 0.544$). The portfolio exhibits a defensive profile (market beta ≈ 0.58) and a descriptive Sharpe ratio superior to the Ibovespa, although the difference is not statistically significant under the Jobson-Korkie test.

The paper's contributions are twofold. The first is to demonstrate that the three-factor model is insufficient to fully explain the returns of the analyzed hybrid portfolio — there exists a component of genuine alpha not captured by the investigated systematic factors. The second is to show that the non-significance of HML for a portfolio with explicit value criteria suggests that factor exposure structure in the Brazilian market can diverge substantially from the international pattern when quality criteria override value criteria — a finding with relevant implications for the domestic finance research agenda.

## 5.2 Limitations

This study presents relevant limitations that must be considered when interpreting the results.

**Small sample and survivorship bias:** the portfolio consists of only 13 assets. The absence of companies that satisfied the filter at the time of selection but subsequently went bankrupt, were delisted, or suffered severe deterioration introduces survivorship bias and tends to systematically overstate results. Generalizing the findings to other fundamentalist portfolios requires caution.

**Absence of rebalancing and residual look-ahead bias:** the fundamentalist screen was applied only once. In a real implementation, the manager would periodically update the selection based on the most recent financial statements. The absence of rebalancing implies that, beyond a certain point, portfolio assets may no longer meet the screen's criteria. Additionally, the effective availability of fundamental data at the start of 2000 for all assets was not verified in detail, which may introduce residual look-ahead bias.

**Data snooping and potential technical overfitting:** the SMA 9-46 parameterization was selected based on Wolwacz (2010), but this does not fully eliminate the risk of data snooping. The sensitivity tests with alternative configurations mitigate, but do not eliminate, this concern. Sullivan, Timmermann, and White (1999) argue that only universe data snooping tests provide complete control.

**Absence of transaction costs:** the simulation does not incorporate brokerage fees, exchange fees, taxes, or bid-ask spreads. Technical timing strategies generate a high number of operations, making the impact of these costs potentially significant. Studies of technical trading in emerging markets suggest total costs of 0.5% to 1.0% per operation at B3, which could materially reduce the identified alpha on a net basis.

**Aggregation method and correlation effects:** the reported results represent the simple average of the 13 assets, not the return of a portfolio with dynamic allocation or capitalization weighting. This methodology understates the importance of inter-asset correlation effects in total risk management.

**Limitations of factor models:** the $R^2$ of the estimated models (0.23 to 0.28) indicates that a substantial fraction of return variance is not explained by the investigated factors. The Fama and French (2015) five-factor model or the Hou, Xue, and Zhang (2015) q-factor model could provide additional explanatory power, capturing dimensions related to quality, profitability, and investment.

### 5.3 Future Research Agenda

Based on the limitations identified and the study's findings, the priority future research agenda includes: (i) the investigation of the non-significance of HML in broader samples of fundamentalist portfolios in Brazil, to determine whether this finding is robust or specific to this configuration; (ii) the application of the Fama and French (2015) five-factor model — incorporating profitability (RMW) and investment (CMA) — which may capture dimensions of the return related to the quality profile of the selected companies; (iii) estimation with periodic rebalancing of the fundamentalist screen and incorporation of realistic transaction costs, bringing the backtesting closer to a practical implementation; and (iv) testing the Hou, Xue, and Zhang (2015) q-factor model as an alternative to the Fama-French framework. The frontier between genuinely differentiated active management and mere replicable factor exposure remains a central question in contemporary financial science — and especially urgent for a market like Brazil, undergoing an accelerating process of institutional maturation and growing competition for capital.

## Declarations

**Conflicts of Interest:** *The authors declare no relevant conflicts of interest related to this article.*

**Funding:** *This study received no specific funding from agencies in the public, private, or non-profit sectors.*

**Data and Code Availability:** *The data used in this study are publicly available from the sources listed in Section 3 (NEFIN/FEA-USP, BCB, Yahoo Finance). The Python scripts used for the analyses are available upon request to the authors.*